\documentclass{aa1}
\usepackage{graphicx}
\usepackage[varg]{txfonts}
\usepackage{cases}
\usepackage{natbib}

\begin{document}

\title{Localizing short-period pulsations in hard X-rays and $\gamma$-rays during an X9.0 flare}

\author{Dong~Li\inst{1}}

\institute{Purple Mountain Observatory, Chinese Academy of Sciences, Nanjing 210023, China \email{lidong@pmo.ac.cn} \\}

\date{Received; accepted}

\titlerunning{Localizing short-period pulsations in HXRs and $\gamma$-rays during an X9.0 flare}
\authorrunning{Dong Li}

\abstract {The feature of Quasi-periodic pulsations (QPPs) is
frequently observed in the light curve of solar/stellar flares.
However, the short-period QPP is rarely reported in the high energy
range of hard X-rays (HXRs) and $\gamma$-rays.} {We investigated the
QPP at a shorter period of about 1~s in high-energy channels of HXRs
and $\gamma$-ray continuum during an X9.0 flare on 2024 October 03
(SOL2024-10-03T12:08).} {The X9.0 flare was simultaneously measured
by the Hard X-ray Imager (HXI), the Konus-Wind (KW), and the
Spectrometer/Telescope for Imaging X-rays (STIX). The shorter period
was determined by the fast Fourier transform with a
Bayesian-based Markov Chain Monte Carlo and the wavelet analysis
method. The HXR images were restructured from HXI and STIX
observations.} {The flare QPP at a shorter period of about 1~s was
simultaneously observed in HXI~20$-$50~keV, 50$-$80~keV and
80$-$300~keV, and KW~20$-$80~keV, 80$-$300~keV, and 300$-$1200~keV
during the impulsive phase of the white-light flare. The
restructured images show that the HXR sources are mainly separated
into two fragments, regarding as double footpoints. Moreover, the
footpoints move significantly during the flare QPP. Our results
suggest that the intermittent and impulsive energy releases during
the powerful flare are mainly caused by the interaction of hot plasma
loops that are rooted in double footpoints.} {We localized the flare QPP
at a shorter period of about 1~s in HXR and $\gamma$-ray continuum
emissions during a white-light flare, which is well explained by the
interacting loop model.}

\keywords{Sun: white-light flares ---Sun: oscillations --- Sun:
X-rays, gamma-rays --- magnetic reconnection}

\maketitle

\section{Introduction}
Quasi-periodic pulsations (QPPs) are common phenomena that are
strongly variable modulations of flare emissions, which are often
characterized by a number of successive, impulsive, and repetitive
pulsations in time-dependent intensity curves during solar/stellar
flares \citep[e.g.,][for a recent reference]{Zimovets21}. A typical
flare QPP usually takes abundant features of temporal
characteristics and plasma radiation of the flare core, and thus it
plays a crucial role in diagnosing coronal parameters and energy
releases on the Sun or Sun-like stars \citep{Yuan19,Inglis23,Li24a}.
The flare QPP was first noted by \cite{Parks69} in wavebands of hard
X-ray (HXR) and microwave. Since then, it has been detected throughout
the electromagnetic spectrum, i.e., in the wavelength range of
radio or microwave, visible, H$\alpha$, Ly$\alpha$, ultraviolet
(UV), extreme ultraviolet (EUV), soft or hard X-ray (SXR/HXR), and
even $\gamma$-rays
\citep[e.g.,][]{Nakariakov10,Tan12,Ning17,Dominique18,Li20,Li24b,Knuth20,Shen22,Zimovets22,Huang24,Karlicky24,Zhou24}.
The flare QPP is commonly a multi-waveband behavior, for instance,
it manifests similarly across a broad range of wavebands, and this is
mainly due to the abundant observational data
\citep{Li15,Li21,Clarke21}. The study of flare QPPs is crucial,
since these could be regarded as a signature of the fundamental physical
process that occurs in solar flares, which might be highly
associated with the intermittent magnetic reconnection, repetitive
particle accelerations, and magnetohydrodynamic (MHD) waves
\citep{Zimovets21,Inglis23}.

The term `QPP' refers to the flare time series consisting of at
least three or four successive pulsations
\citep{McLaughlin18,Inglis24}, although this may not be a strict
definition. The term `period' is expected to be stationary,
that is, the lifetimes of all pulsations for one QPP should be
equivalent. However, the observed QPPs are often non-stationary,
i.e., with a varying instantaneous period, regarded as a
`quasi-period' \citep{Nakariakov19}. In terms of observations,
flare QPPs have been reported over a broad timescale, and the
magnitude order of quasi-periods can range from milliseconds through
seconds to minutes
\citep[e.g.,][]{Tan10,Brosius18,Carley19,Kashapova21,Li22a,Li22,Collier23,Zhao23,Inglis24}.
Flare QPPs at the period of sub-seconds are often observed in
the radio emission \citep{Tan10,Yu19,Karlicky24}, mainly
because of the high cadence that can be reached in the radio band.
Conversely, the detection of flare QPPs at X-ray energies is often
on a timescale of seconds and minutes, i.e., $\geq$4~s
\citep{Tan16,Hayes20,Collier23}, largely due to the observational
constraints. Three spiking intervals were identified with Fermi
Gamma-ray Burst Monitor (GBM) data, and only one was found to show a
periodicity at the frequency of 1.7$\pm$0.1~Hz
\cite[cf.][]{Knuth20}. By systematically analyzing solar flares
recorded by Fermi/GBM in the burst mode, \cite{Inglis24} conclude
that the QPPs with periods shorter than 5~s have a low base
occurrence rate. Moreover, Fermi/GBM cannot localize the X-ray
source region. In this letter, we localized a flare QPP at the
period of about 1~s in HXR and $\gamma$-ray continuum emissions
during a powerful flare.

\section{Observations}
We analyzed an X9.0 flare that occurred on 2024 October 03,
and that was situated in the active region of NOAA~13842. The flare was
simultaneously measured by the Hard X-ray Imager
\citep[HXI;][]{Su19} and the Ly$\alpha$ Solar Telescope
\citep[LST;][]{Feng19} for the Advanced Space-based Solar
Observatory\citep[ASO-S;][]{Gan19}, the Konus-Wind
\citep[KW;][]{Lysenko22}, the Geostationary Operational
Environmental Satellite (GOES), the Spectrometer/Telescope for
Imaging X-rays \citep[STIX;][]{Krucker20} on board the Solar
Orbiter, the Solar Upper Transition Region Imager
\citep[SUTRI;][]{Bai23}, and the Chinese H$\alpha$ Solar
Explorer\citep[CHASE;][]{Lic19}.

HXI is used to image the solar flare in HXR channels of about
10$-$300~keV. The time cadence is as high as 0.25~s in burst
mode. KW is used to investigate $\gamma$-ray bursts and solar
flares. The count rate light curves have varying time cadences
(e.g., 0.002$-$0.256~s) in flare mode. STIX can provide
flare imaging spectroscopy in the energy range of 4$-$150~keV at a
time cadence of about 1~s. GOES records the solar SXR
radiation in channels of 1$-$8~{\AA} and 0.5$-$4~{\AA} at a time
cadence of 1~s.

CHASE takes spectroscopic observations of the full-disk Sun in
passbands of H$\alpha$ and Fe~I. The spatial scale is about
1.04\arcsec\ per pixel, and the time cadence is about 71~s. The
Solar Disk Imager (SDI) for LST captures the full-disk map at
Ly$\alpha$~1216~{\AA}, the time cadence is normally 60~s. The
White-light Solar Telescope (WST) for LST takes the white-light
snapshot at 3600~{\AA}, the normal time cadence is 120~s. SUTRI
captures a full-disk map at a temperature of about 0.5~MK, and it
uses the Ne VII 465~{\AA} line \citep{Tian17}. The time cadence is
about 31~s, and the spatial scale is about 1.23\arcsec\ per pixel.

\section{Methods and Results}
\begin{figure}
\centering
\includegraphics[width=0.9\linewidth,clip=]{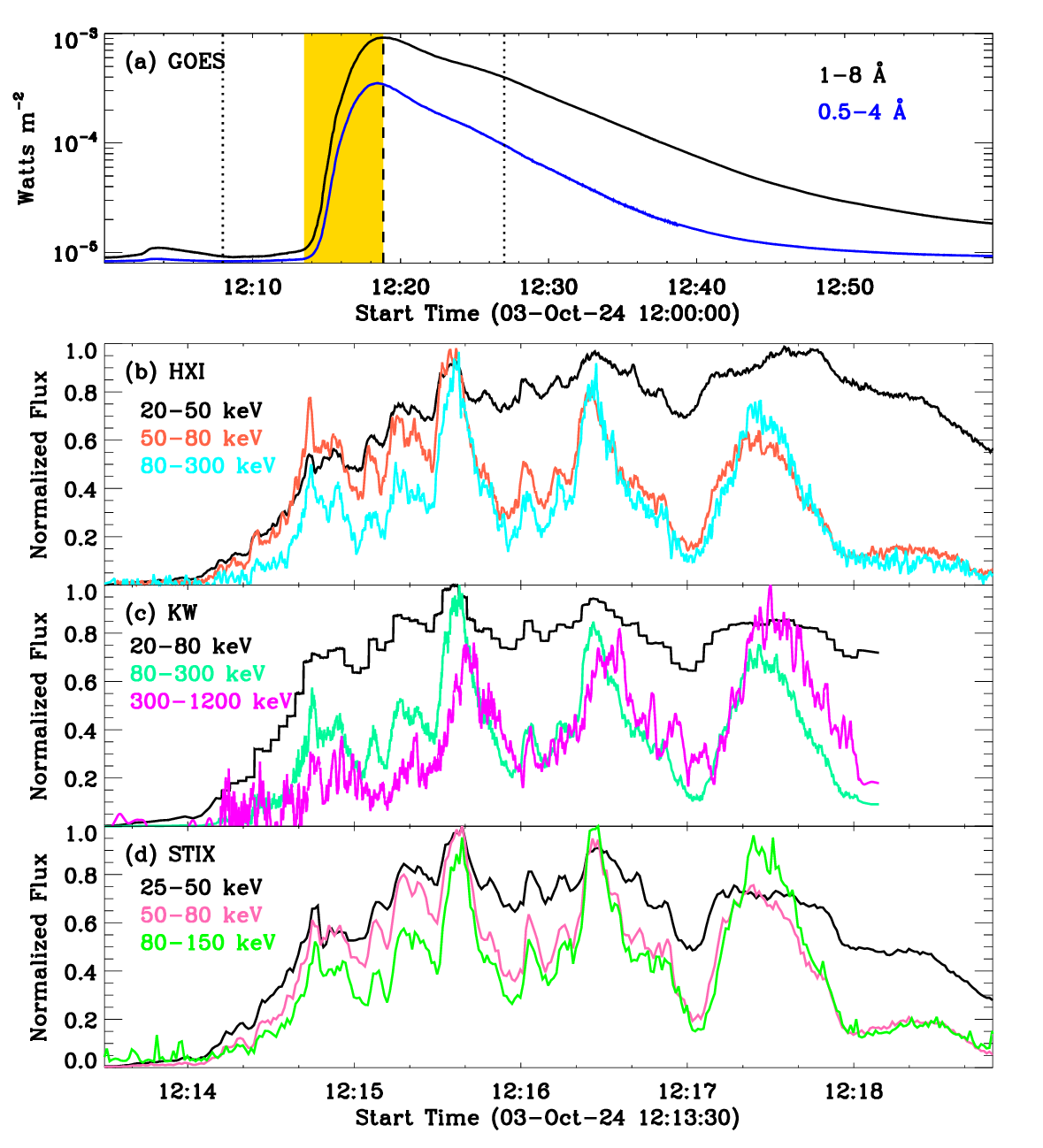}
\caption{Light curves of the solar flare on 2024 October 03. (a):
Full-disk light curves during 12:00$-$13:00~UT measured by GOES in
wavelengths of 1$-$8~{\AA} (black) and 0.5$-$4~{\AA} (blue). The
vertical lines mark the start, peak, and stop times of the X9.0
flare. (b): Light curves from 12:13:30~UT to 12:18:50~UT measured by
ASO-S/HXI in the energy range of 20$-$50~keV (black), 50$-$80~keV
(tomato), and 80$-$300~keV (cyan). (c): Light curves between
12:13:30~UT and 12:18:50~UT recorded by KW in channels of
20$-$80~keV (black), 80$-$300~keV (spring green), and 300$-$1200~keV
(magenta). (c): Light curves between 12:13:30~UT and 12:18:50~UT
observed by STIX in channels of 20$-$50~keV (black), 50$-$80~keV
(hot pink), and 80$-$150~keV (green). \label{flux}}
\end{figure}

Figure~\ref{flux} shows the light curves in multiple wavebands
during the powerful flare on 2024 October 03. The SXR flux recorded
by GOES~1$-$8~{\AA} suggests an X9.0 flare, which begins at about
12:08:00~UT, peaks at about 12:18:50~UT, and stops at about
12:27:00~UT, as marked by the vertical lines in panel~(a).
Figure~\ref{flux}~(b)$-$(c) shows the light curves in the
high-energy range of HXRs (20$-$300~keV) and $\gamma$-ray continuum
(300$-$1200~keV) measured by HXI, KW and STIX during
12:13:30$-$12:18:50~UT, as outlined by the gold shadow in panel~(a).
The HXI fluxes were derived from an open flux monitor and have the
highest time cadence of 0.25~s. KW fluxes have been interpolated
into a uniform time cadence of 0.256~s, since the raw light curves
have a varying time cadence. The STIX fluxes were extracted from the
pixelated science data, which have a time cadence of 1~s. They all
reveal some successive pulsations with a large amplitude, which
could be regarded as flare QPPs. These large-amplitude QPPs appear
to have longer quasi-periods, i.e., $>$10~s. On the other
hand, there are many repeated and successive wiggles that are
superimposed on the large-amplitude pulsations, which could be
regarded as small-amplitude oscillations at a shorter
period, termed short-period QPPs. The short-period QPPs can be
clearly seen in the light curves measured by HXI and KW, but they
were not observed by STIX due to its low time cadence.

In order to identify the shorter period, a fast Fourier Transform
(FFT) was applied for the raw light curves with the Lomb-Scargle
periodogram method \citep{Scargle82}, and the Fourier power spectral
density (PSD) was obtained. Then, the Bayesian-based Markov Chain
Monte Carlo (MCMC) approach was utilized to fit the PSD with a simple
model ($M$) that consists of a power-law distribution and a
constant ($C$) term
\citep[cf.][]{Liang20,Anfinogentov21,Guo23,Shi23,Inglis24}, as shown
in Eq.~(\ref{eq1}):
\begin{equation}
\centering
 M (f) = A f^{\alpha} + C.
\label{eq1}
\end{equation}
\noindent Here, $f$ denotes to the Fourier frequency, $A$ is the
amplitude, $\alpha$ is the power-law index. The MCMC-fit results for
the observational data were determined by this simple model.

\begin{figure}
\centering
\includegraphics[width=0.9\linewidth,clip=]{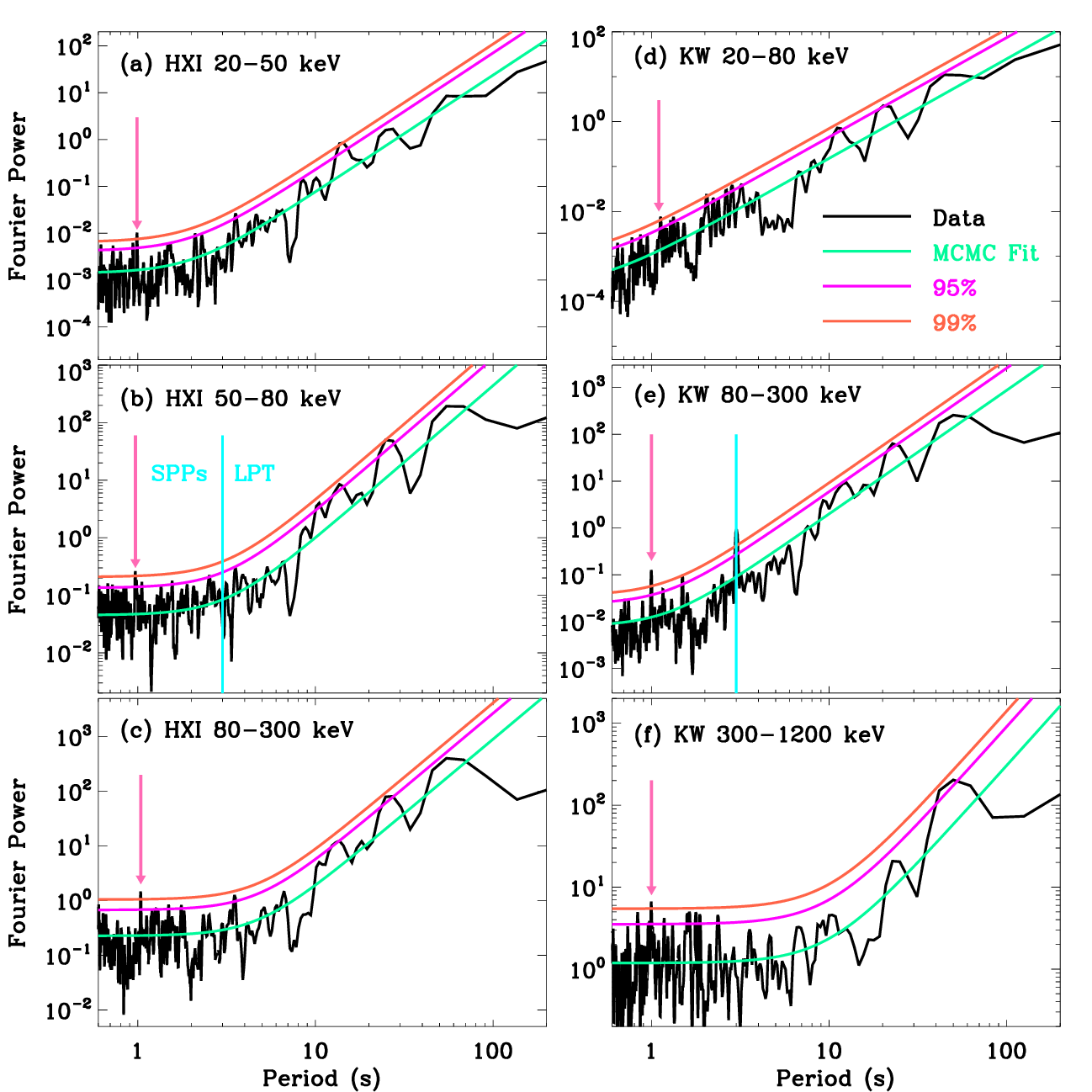}
\caption{Fourier PSDs and their MCMC-fit results in log-log space.
The spring green line in each panel indicates the MCMC fit for the
observational data (black), the magenta and tomato lines represent
the confidence levels at 95\% and 99\%, respectively. The hot pink
arrow outlines the interested period above the 99\% confidence level.
The cyan vertical line divides the spectrum into SPPs and LPT
components. \label{fft1}}
\end{figure}

Figure~\ref{fft1} presents the Fourier PSDs and their MCMC-fit
results in high-energy channels measured by HXI and KW. We note that
several quasi-periods exceed the 95\% confidence level in both HXRs
and $\gamma$-ray continuum emissions, including the large- and
small-amplitude QPPs, which correspond to the longer periods that
are bigger than 10~s and the shorter period at about 1~s. On the
other hand, one quasi-period at about 1~s is above the 99\%
confidence level, confirming that the short-period QPP real
exists in HXRs and the $\gamma$-ray continuum, as indicated by the
hot pink arrow. It should be pointed out that the 3-s period in KW
fluxes is attributed to the Wind's rotation, resulting in
millisecond timescales as dips in the light curves at a period of
3~s \citep[cf.][]{Lysenko22}.

To look closely at the short-period QPP, the wavelet analysis method
with a mother function of `Morlet' \citep{Torrence98} is applied to
the detrended time series that contains the short-period pulsations
(SPPs). Using the FFT method with a Gaussian filter function
\citep{Li17,Ning17,Shi23}, the raw light curve in each channel was
decomposed into two components: the long-period trend (LPT) and the
SPP. Here, we used a time threshold of 3~s to distinguish the two
components, as indicated by the cyan vertical line in
Figure~\ref{fft1}. This time threshold can separate the LPT and
SPPs components well, because both the longer (i.e., $>$10~s) and shorter
periods ($\sim$1~s) are far away from it.

\begin{figure}
\centering
\includegraphics[width=0.9\linewidth,clip=]{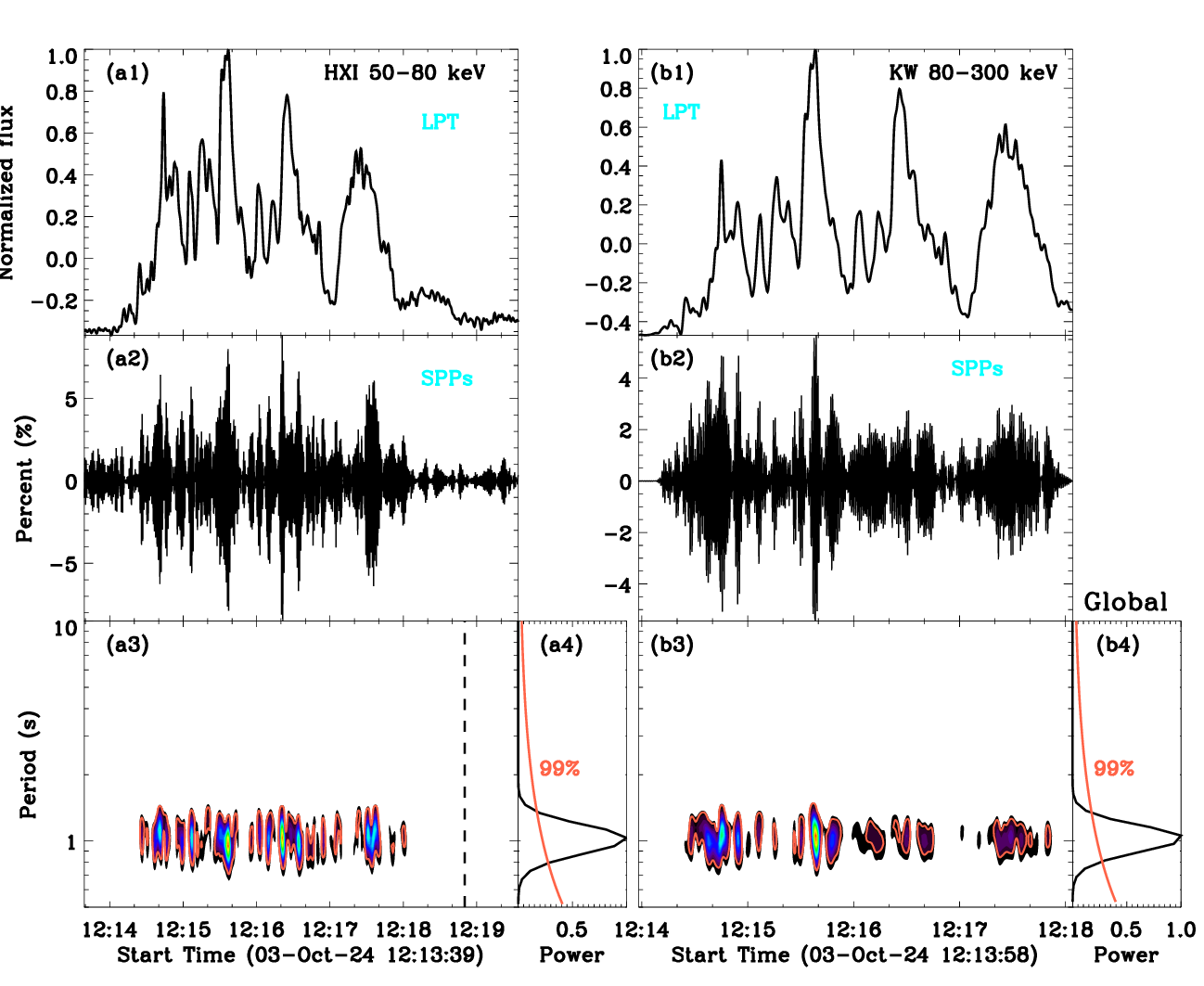}
\caption{Morlet wavelet analysis results. (a1-b2): Long-period
trend (LPT) and short-period pulsations (SPPs) derived from raw light
curves with the FFT filter method. They have been normalized by the
maximum value of the LPT component. (a3 \& b3): Morlet wavelet power
spectra. (a4 \& b4) Global wavelet power spectra. The tomato
contours and lines indicate the significance level of 99\%.
\label{wav}}
\end{figure}

Figure~\ref{wav} shows the wavelet analysis results in HXR channels
of HXI~50$-$80~keV (a1-a4) and KW~80$-$300~keV (b1-b4). Panels~(a1)
and (b1) show the normalized time series of the LPT component with the
FFT filter method, and they have been normalized by their maximum
intensities. They both reveal QPP signals at longer periods, i.e.,
$>$10~s. Thus, the LPT component could also be regarded as the
strong background. Panels~(a2) and (b2) draw the time series of SPPs
components. They both show the QPP patterns at a shorter period. The
modulation depth of the short-period QPP, which is defined as the
ratio between the SPP component and the maximum intensity of LPT
component, is much less than 10\% at both HXI~50$-$80~keV and
KW~80$-$300~keV. The averaged modulation depth for the short-period
QPP is estimated to about 5\%. Figure~\ref{wav}~(a3)$-$(b4) show the
wavelet power spectra and Global wavelet power spectra for the SPP
component. They are dominated by a bulk of power spectra inside the
99\% significance level, and all these power spectra are centered at
about 1~s, which is consistent with the FFT power spectra. Moreover,
the 1-s period tends to appear in the peaks of the LPT component
during the flare impulsive phase, i.e., from about 12:14~UT to
12:18~UT. Lastly, we present the cross-correlation
analysis between the QPP patterns detected in different
instruments, i.e., ASO-S/HXI and KW, as shown in Figure~\ref{corr}.
The linear Pearson correlation coefficient has a maximum value of
about 0.46, which occurs at the time lag of zero, as marked by the
vertical line. The correlation analysis indicates that there are not
phase shifts between the QPP signals.

\begin{figure}
\centering
\includegraphics[width=0.9\linewidth,clip=]{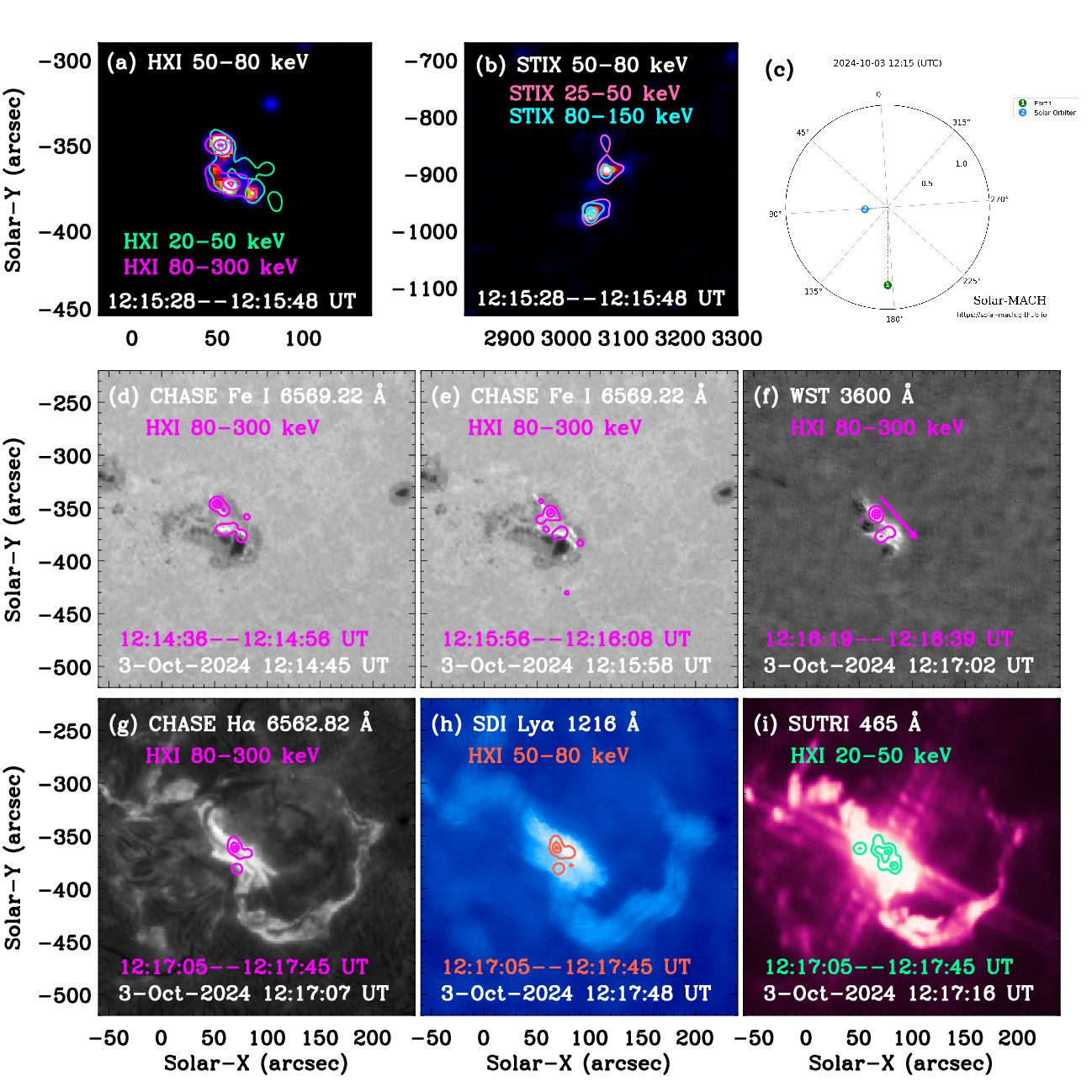}
\caption{Multi-wavelength images during the X9.0 flare. (a \& b):
HXR maps restructured from the HXI and STIX observational data. (c)
Sketch plot of the spatial location of STIX and its connection with
the Sun and Earth at 12:15~UT on 2024 October 03. (d-f) White-light
sub-maps measured by CHASE~Fe~I and WST~3600~{\AA}. The magenta
arrow indicates the movement direction of HXR sources. (g \& h):
H$\alpha$ and Ly$\alpha$ images observed by CHASE and SDI. (i): The
EUV snapshot captured by SUTRI at 465~{\AA}. The contour levels are
set at 10\%, 50\%, and 90\%, respectively. \label{smap}}
\end{figure}

Figure~\ref{smap} presents the spatial structure of the QPP
pattern in multiple wavebands during the X9.0 flare. Panels~(a) and
(b) show the HXR maps in the energy range of 50$-$80~keV, and the
color contours represent HXR radiation in other energy ranges. In
this study, the HXR map from HXI was reconstructed by the HXI\_CLEAN
algorithm with a pixel scale of 2\arcsec. The HXR map from STIX was
reconstructed from the expectation-maximization (EM) algorithm in
the STIX Aspect System \citep{Warmuth20}. Utilizing the Solar-MACH
\citep{Gieseler23}, panel~(c) plots the spatial location of STIX and
its connection with the Sun and Earth at 12:15~UT on 2024 October
03. In this event, STIX's in situ measurements, 84.9 degrees east of
the Sun-Earth line at 0.299~AU, provided a unique vantage point,
along with the Earth measurements at 1~AU.Both HXI and STIX
maps show two main HXR sources in the energy range of 50$-$300~keV
or 50$-$150~keV, which could be considered as conjugate
points that connected by hot flare loops, as indicated by the X-ray
emission at HXI~20$-$50~keV (green contours). Assuming that the
flare loop has a semi-circular profile \citep[cf.][]{Tian16,Li22},
the flare loop length ($L$) can be estimated by the distance between
the conjugate points, which is roughly 50~Mm. The minor
radius ($r$) of the flare loop can be estimated by the conjugate
points when assuming a circular shape for the flare loop, and it is
about 2.5~Mm.

Figure~\ref{smap}~(d)$-$(f) show the white-light images measured by
CHASE~Fe~I and ASO-S/WST~3600~{\AA} at three times, and the overlaid
contours are HXR radiation at HXI~80$-$300~keV. Here, the
running-difference images from ASO-S/WST~3600~{\AA} is shown to
highlight the white-light emission. The three images reveal two
bright patches at the edge of a sunspot group, suggesting that the
X9.0 flare is a white-light flare. Moreover, the white-light
brightening area matches the HXR radiation sources, indicating that
the white-light emission is strongly associated with the nonthermal
radiation. The HXR sources move significantly toward the south-west
direction, as indicated by the magenta arrow in panel~(f). The
displacement ($D$) of these two motions can be determined by the
distance between the brightness centers of two adjacent footpoints,
which are roughly 24~Mm and 16~Mm, respectively. Panels~(g) and (h)
present the H$\alpha$ and Ly$\alpha$ images observed by CHASE and
ASO-S/LST, respectively. The flare radiation in wavebands of
H$\alpha$ and Ly$\alpha$ is mainly from the upper chromosphere and
low transition region on the Sun. Two strong ribbon-like features
that match double HXR sources (magenta contours) can be clearly seen
in the H$\alpha$ and Ly$\alpha$ images, suggesting that the two
strong ribbon-like features are double ribbons of the flare. On the
other hand, a slender and elongated ribbon-like structure can also
be seen in H$\alpha$ and Ly$\alpha$ images, and it appears to have a
semi-circular shape. It is much weaker than the double ribbons, and
it is not covered by the HXR emission. Figure~\ref{smap}~(i) shows
the EUV image captured by SUTRI at 465~{\AA}, which is mainly formed
in the upper transition region or the low corona. We can find that a
bright loop-like structure appears in the EUV image, and it connects
the double ribbons in H$\alpha$ and Ly$\alpha$ images. A faint
semi-circular structure can also be seen in the SUTRI~465~{\AA}
image, which overlaps with the slender and elongated ribbon in
the upper chromosphere or the low transition region.

\section{Discussions}
We systematically analyzed an X9.0 flare on 2024 October 03 that was
simultaneously observed by ASO-S/HXI, KW, STIX, SUTRI, CHASE, and
ASO-S/LST in wavebands of HXR, $\gamma$-ray continuum, EUV,
Ly$\alpha$, H$\alpha$, and white light. The X9.0 flare shows
significant enhancements in wavebands of WST~3600~{\AA} and CHASE
Fe~I~6569.22~{\AA}, indicating a white-light flare. The HXI and KW
light curves in the high-cadence burst mode provide us with an
opportunity to investigate the short-period QPP in the energy range
of HXRs and $\gamma$-ray continuum. Using the FFT method with a
Bayesian-based MCMC approach \citep{Anfinogentov21,Guo23,Shi23}, a
quasi-period centered at about 1~s was simultaneously identified in
channels of HXI~20$-$50~keV, 50$-$80~keV and 80$-$300~keV,
KW~20$-$80~keV, 80$-$300~keV and 300$-$1200~keV. The shorter period was
also determined by the wavelet analysis method, and it appears to
enhance during the impulsive phase of the X9.0 flare, especially in
the HXR pulse time. The modulation depth of the short-period QPP,
which was determined by the ratio between the SPPs and its LPT, was
estimated to about 5\% in average, suggesting that the 1-s period
is a weak QPP signal. This is different from that of the long-period
QPPs in HXRs and $\gamma$-rays, which often have a large modulation
depth \citep[e.g.,][]{Nakariakov10,Li22,Li24c}.

The flare QPPs at the very short period, i.e., $\ll$1~s, are
frequently observed in wavebands of radio and microwave emissions
\citep{Tan10,Yu19,Karlicky24}, which is attributed to the higher
time resolution and the higher signal-to-noise ratio of solar radio
telescopes. On the contrary, the flare QPPs in the high-energy range
of HXRs and $\gamma$-rays are often identified to have a
characteristic period that exceeds 4~s
\citep[e,g.,][]{Parks69,Nakariakov10,Li22,Collier23,Inglis24},
possibly due to the observational limitation, that is, the
typical signal-to-noise ratio for the HXR instrument is lower. By
using the Fermi/GBM data in the burst mode, a solar flare was found
to show the periodicity with a frequency of 1.7$\pm$0.1~Hz
\citep[cf.][]{Knuth20}, demonstrating the presence of shorter
periods in the HXR channel. On the other hand, a statistical study
based on the Fermi/GBM data \citep[e.g.,][]{Inglis24} suggests that
the short-period QPPs at HXRs are not widespread, although they
identified a few shorter periods at the timescale of about 1$-$4~s,
similar to our results. In our case, the shorter period centered at
about 1~s is also detected in the $\gamma$-ray continuum, and the
QPP sources are localized by using the HXI and STIX data in two
different views, which can make us to explore its generation
mechanism.

The generation mechanism of flare QPPs is still an open issue
\citep{Zimovets21,Inglis23}. Here, we discussed the possible
generation mechanism of the short-period QPP. The flare QPPs are
often associated with the MHD wave. In our case, the phase speed
($c_P$) can be estimated from the loop length ($L$) and the period
($P$), such as $c_P=2L/P \approx 1.0 \times 10^5$~km~s$^{-1}$. The
phase speed is much faster than the sound and Alfv\'{e}n velocities
of local plasmas. Therefore, the short-period at about 1~s might be
modulated by a fast-mode wave, such as the global sausage wave. The
fast kink wave is impossible, because it is essentially
compressive, but it becomes `weakly compressive' or `almost
incompressive' in the long-wavelength limit. On the other hand, the
global sausage wave in the solar corona requires that the plasma loop
must be sufficient thick and dense \citep{Nakariakov03}, that is,
the density ratio $\rho_{io}$ inside and outside the flare loop
should be satisfied with Eq.~\ref{eq2}:

\begin{equation}
\centering
\rho_{io} \gg (\frac{L}{1.3 r})^{2}.
\label{eq2}
\end{equation}
\noindent In our case, the density ratio is at least 237, if the
shorter period is modulated by the global sausage wave. Such a ratio
is much larger than the one detected in flare loops
\citep[e.g.,][]{Tian16}. So, it is difficult for the short-period
to be modulated by the global sausage wave. We note that the
flare footpoints evidently move, indicating that the flare loop
also moved. Thus, the lateral separations (D) of the flare loop with
the variation of time can be estimated to about 24~Mm and 16~Mm,
respectively. The time differences ($\tau$) were determined by the
center times in Figure~\ref{smap}~(d)$-$(f), which are 76~s and 27~s.
At last, the Alfv\'{e}n speed ($v_A$) should be of the order
\citep{Emslie81}:

\begin{equation}
\centering
v_{A} \simeq \frac{D}{\tau}.
\label{eq3}
\end{equation}
\noindent Here, the Alfv\'{e}n speed is estimated to about
300$-$600~km~s$^{-1}$, which agrees with the pervious result
\citep{Emslie81}. Therefore, the short-period QPP may be explained
by the interacting loop model presented by \cite{Emslie81}. In this
model, the small pulsations at short periods are regarded as the
successive activations of a series of hot plasma loops. That is, the
unstable hot loop induces a HXR pulse and then leads to the lateral
expansion of magnetic lines to the nearby plasma loop, resulting
in the instability and disturbance of the neighboring loop. In
this process, the nonthermal electrons can be rapidly
accelerated, and such process will continue to occur from one hot
loop to the neighboring loop, generating the successive and repeated
HXR pulsations during the X9.0 flare \citep[cf.][]{Zhao23}. It
should be pointed out that the interaction of different hot loops
was not detected, mainly due to the observational limitation, i.e.,
the insufficient spatial resolution and the low signal-to-noise
ratio. Previous studies have suggested that the flare loop
essentially consists of a number of fine-scale hot plasma loops, and
they are usually regarded as a loop system
\citep[e.g.,][]{Tian16,Li23}, which may be attributed to the diffuse
nature of the EUV/SXR radiation. In a word, the interacting loop
model can easily be used to explain the short-period QPP.

\section{Summary}
Combining the observational data measured by ASO-S/HXI, KW, STIX,
CHASE, SUTRI, and GOES, we investigated the short-period QPP during
a white-light flare. Our main conclusions are summarized as
follows:

(1) The short-period pulsations are simultaneously seen in the
high-energy range of HXRs and the $\gamma$-ray continuum. Using the FFT
and the wavelet analysis method, the quasi-period was measured to
about 1~s .

(2) The modulation depth, which was determined by the ratio between
the SPPs and its LPT, was estimated to about 5\% on average,
indicating a weak QPP signal.

(3) The short-period QPP during the flare impulsive phase can be
interpreted by the interacting loop model presented by \cite{Emslie81}.

\begin{acknowledgements}
The author would like to thank the referee for his/her
inspiring comments. This work is supported by the Strategic
Priority Research Program of the Chinese Academy of Sciences, Grant
No. XDB0560000, the National Key R\&D Program of China
2022YFF0503002 (2022YFF0503000). We thank the teams of ASO-S/HXI,
STIX, KW, CHASE, SUTRI, and GOES for their open data use policy.
ASO-S mission is supported by the Strategic Priority Research
Program on Space Science, the Chinese Academy of Sciences, Grant No.
XDA15320000. SUTRI is a collaborative project conducted by the
National Astronomical Observatories of CAS, Peking University,
Tongji University, Xi'an Institute of Optics and Precision Mechanics
of CAS and the Innovation Academy for Microsatellites of CAS. The
CHASE mission is supported by China National Space Administration
(CNSA). The STIX instrument is an international collaboration
between Switzerland, Poland, France, Czech Republic, Germany,
Austria, Ireland, and Italy.
\end{acknowledgements}

\begin{appendix}

\section{Cross-correlation analysis}
\begin{figure}[h]
\centering
\includegraphics[width=0.9\linewidth,clip=]{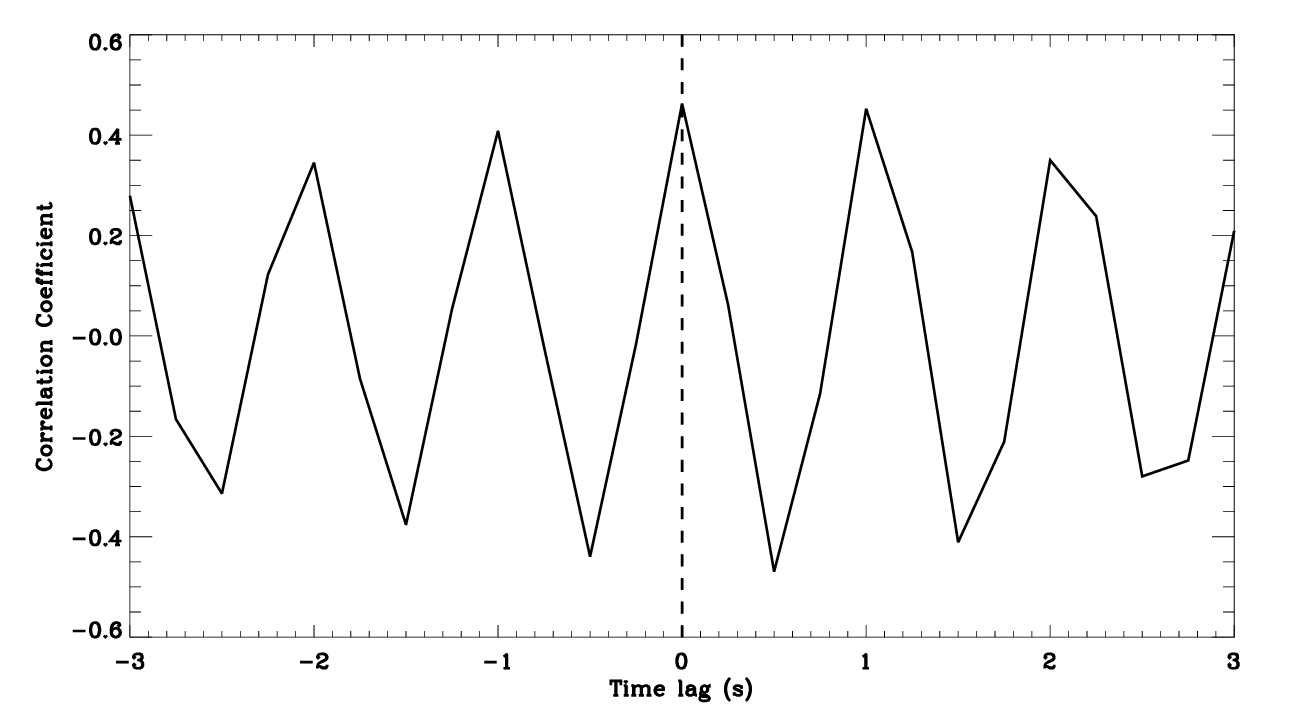}
\caption{Cross-correlation analysis between the QPP patterns
detected in two different instruments, i.e., HXI and KW. The time
series represents the cross-correlation coefficients as a function
of the time lag. The vertical line marks the maximum coefficient.
\label{corr}}
\end{figure}

\end{appendix}

\end{document}